\documentclass[sigconf]{acmart}
\usepackage[utf8]{inputenc}
\pdfoutput=1
\usepackage{booktabs} % For formal tables
\usepackage{xspace}
\usepackage{tikz}
\usetikzlibrary{calc}
\newcommand{\xlap}{\textsc{X-Lap}\xspace}
\setcopyright{rightsretained}
\acmDOI{https://doi.org/10.1145/3267419.3267422}

\acmISBN{}

\acmConference[ECRTS-RTN'17]{RTN'2017 The 15th International Workshop on Real-Time Networks}{June 2017}{Dubrovnik, Croatia}
\acmYear{2017}
\copyrightyear{2017}

\acmPrice{}

\begin{document}
\title[\xlap: A Systems Approach for Cross-Layer Profiling and\\ Latency Analysis for Cyber-Physical Networks]{\xlap: A Systems Approach for Cross-Layer Profiling and Latency Analysis for Cyber-Physical Networks}

\author{Stefan Reif}
\affiliation{\institution{Friedrich-Alexander University Erlangen-N\"urnberg}}
\email{reif@cs.fau.de}

\author{Andreas Schmidt}
\affiliation{\institution{Saarland Informatics Campus}}
\email{andreas.schmidt@cs.uni-saarland.de}

\author{Timo Hönig}
\affiliation{\institution{Friedrich-Alexander University Erlangen-N\"urnberg}}

\email{thoenig@cs.fau.de}

\author{Thorsten Herfet}
\affiliation{\institution{Saarland Informatics Campus}}
\email{herfet@cs.uni-saarland.de}

\author{Wolfgang Schröder-Preikschat}
\affiliation{\institution{Friedrich-Alexander University Erlangen-N\"urnberg}}
\email{wosch@cs.fau.de}

% The default list of authors is too long for headers}
\renewcommand{\shortauthors}{S. Reif et al.}

\begin{abstract}
Networked control applications for cyber-physical networks demand predictable and reliable real-time communication. Applications of this domain have to cooperate with network protocols, the operating system, and the hardware to improve safety properties and increase resource efficiency. In consequence, a cross-layer approach is necessary for the design and holistic optimisation of cyber-physical systems and networks.
This paper presents \xlap, a cross-layer, inter-host timing analysis tool tailored to the needs of real-time communication. We use \xlap to evaluate the timing behaviour of a reliable real-time communication protocol. Our analysis identifies parts of the protocol which are responsible for unwanted jitter.
% such as IPC.
To system designers, \xlap provides useful support for the design and evaluation of networked real-time systems.
\end{abstract}

%
% The code below should be generated by the tool at
% http://dl.acm.org/ccs.cfm
% Please copy and paste the code instead of the example below.
%


% We no longer use \terms command
%\terms{Theory}


\hyphenation{time-stamp time-stamps cyc-le-stamp cyc-le-stamps}

\maketitle

\section{Introduction}
\label{sec:introduction}

Due to rising interest in novel technologies, for instance in the areas of mobility (i.e. autonomous driving), manufacturing (i.e. smart factories), and augmented environments (i.e. Internet of Things), it is evident that the gap between the digital and physical world is getting narrower. In particular, \emph{Cyber-Physical Systems} (CPSs)~\cite{lee2008cyber} incorporate mechanisms where either world is actively manipulating the other, requiring a holistic view on these systems. This comes with strict requirements regarding latency and resilience of these systems in order to provide efficiency, predictability, and reliability. For every practical CPS application it is necessary to communicate, hence we have to consider \emph{Cyber-Physical Networks} (CPNs) as networks of CPSs, which inherently have the same requirements.

To build efficient and safe interconnected systems, it is imperative to consider a close cooperation of the network infrastructure, the operating system, and the application, treating the systems as a single unit. Only a cross-layer approach ensures the balance of individual components which optimises the system as a whole.

Cross-layer system design and optimisation depend on an appropriate evaluation method. The evaluation has to cover each individual system component, such as the application, the operating system, the protocol stack, and the hardware, but it also has to provide a holistic view onto the system.

There are no existing off-the-shelf approaches for executing an empirical analysis with CPNs. This paper presents \xlap, a system for cross-layer, inter-host timing analysis tailored to the requirements of CPNs. In particular, we use \xlap to evaluate and analyse the real-time communication protocol \emph{Predictably Reliable Real-time Transport} (PRRT), which is further described in Sec.~\ref{subsec:prrt}, in order to identify specific root causes for high latency and unpredictability.

This analysis provides insights to allow tailoring of PRRT to the requirements of CPNs. First, \xlap identifies network, protocol, and operating system bottlenecks regarding timing. The goal is to significantly reduce latency and jitter, optimise resource usage in the network stack and at the communication end-points, and eventually improve quality of control in a CPN. Second, \xlap guides trade-off decisions between network resources and host resources. For instance, \emph{Forward Error Correction} (FEC) is expensive at protocol level, but compensates for network reliability problems. Using a cross-layer approach, \xlap allows to fine-tune FEC parameters to optimise to the specific needs of the application. Third, \xlap can experimentally verify theoretical timing models for CPNs.

The contribution of this paper is threefold:
\begin{itemize}
\item We present \xlap, a cross-layer, inter-host timing analysis tool for real-time networks and CPNs.
\item We evaluate PRRT, a predictably reliable real-time communication protocol, for latency and jitter using \xlap.
\item We identify root causes of timing unpredictability in PRRT.
\end{itemize}

The rest of the paper is structured as follows. First, Sec.~\ref{sec:implementation} introduces the PRRT protocol and discusses the \xlap architecture in detail. In Sec.~\ref{sec:evaluation} we present our evaluation, describe analysis methods, and review our evaluation results. Finally, Sec.~\ref{sec:relatedwork} sets our proposed architecture of \xlap into context with related work, and Sec.~\ref{sec:conclusion} concludes the paper.

\section{Background and Implementation}
\label{sec:implementation}

To determine the validity of our proposed approach we examine and analyse a real-time communication protocol. We use \xlap to analyse the PRRT protocol, which provides predictably reliable real-time communication. \xlap pairs lightweight timing measurement facilities with analysis tools to evaluate latency and jitter.

\subsection{Predictably Reliable Real-time Transport}
\label{subsec:prrt}

\begin{figure*}
    \includegraphics[width=\textwidth]{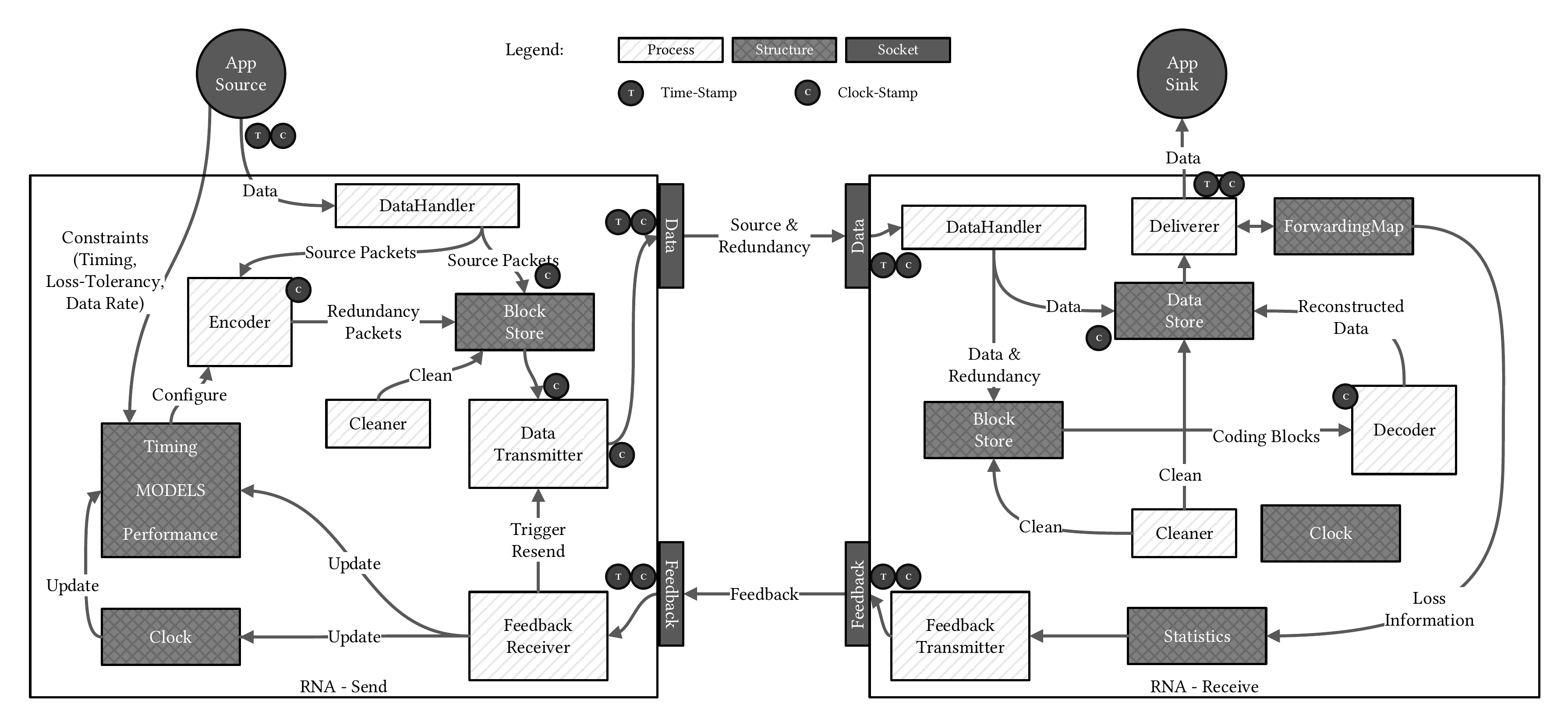}
    \caption{PRRT Architecture and \xlap Integration}
    \label{fig:rna}
    \vspace{-.9em}
\end{figure*}

When it comes to communication protocol support for CPSs, we see a lack of generic approaches that ensure resilience- and latency-awareness during operation. These aspects are important for video broadcast solutions, which is also the origin of our transport layer protocol PRRT~\cite{Gorius:PRRT}. The underlying motivation is the ability of video streaming applications to conceal faults, e.g. by repeating frames, and the requirements towards timely delivery, i.e. past frames can no longer be displayed. Currently, video streaming and processing find their way into CPSs, such as robot control systems, that apply computer vision to monitor their environment. But even without using high data rate streams such as video, there is an inherent fault-tolerance and time-criticality in these systems.

\subsubsection{Error Control and Timeliness}

Consequently, it is necessary to apply appropriate error control approaches, to ensure resilience, but at the cost of increased latency. Fundamental work on the limits of this approach can be found in \cite{Shannon:Math,Polyanskiy:Finite}, defining the relationship between latency and reliability. Choosing an adequate error control strategy requires knowledge about the channel and its evolution over time. Especially in CPS with their low time budgets, we require optimised parameters for error control.

This is achieved by applying a \textit{Hybrid Error Correction}~(HEC) scheme~\cite{Lee:Throughput}, which combines \textit{Automated Repeat reQuest} (ARQ) and FEC into an optimal scheme that is able to approach a channel's capabilities~\cite{Tan:HEC}. Hence, PRRT provides mechanisms to retransmit packets and send redundancy packets along to ensure resilience. For this, it needs to acquire measurements of the channel state, namely loss and delay characteristics. As these characteristics change over time, it requires continuous measurements and adaption of HEC parameters, which leads to an \textit{Adaptive Hybrid Error Correction} (AHEC) scheme. The required adaptivity is provided by the PRRT capability to incorporate application constraints (i.e. throughput, maximum latency) and tolerable residual error, when optimising its coding parameters during operation.

Considering these requirements, it is clear that existing protocols are mostly not suitable for CPNs, which is independent of the network layer they are operating on. Protocols such as TCP provide full-reliability, but no timing guarantees, while lower levels typically do not provide error control. In its current version, PRRT is implemented on the transport layer to be used in IP networks, but the requirements towards the lower layer are minimal. In fact, the same approaches and implementation can run over Ethernet or similar technologies that support basic addressing and forwarding.

\subsubsection{Architecture}
The overall architecture for both sides of the communication is depicted in Fig.~\ref{fig:rna}. Applications interact with the protocol as they would with any other unordered, datagram-oriented communication system. Datagrams are forwarded through the architecture and sent to the channel. Depending on the current coding configuration, blocks of datagrams are grouped and redundancy packets are generated and sent, implementing the FEC part. Upon reception of packets, the receiver forwards the source packets to the application and uses redundancy packets to restore packets that were lost in transit. Feedback to the sender is sent in regular intervals, giving information on the channel and receiver state. The former includes loss and latency readings, while the latter indicates which blocks require additional information packets to allow reconstruction. This feedback procedure implements the reactive ARQ part of error control. Furthermore, the messages are used to synchronise protocol clocks on sender and receiver side, e.g. to clean up packets that have already expired and stop retransmitting or decoding these.

\subsubsection{Interaction with the System Layer}
PRRT provides latency- and resilience-awareness on the protocol layer, enabling advanced applications with error-tolerance and inherent time constraints. Doing so, it relies on underlying layers and can significantly benefit from the reduction of latencies and jitter.

Latency reductions can be achieved using delay hiding, hence executing supporting tasks in a way that the main task does not experience additional latency. In order to achieve this, preparatory and clean-up tasks can be deferred to less busy moments in time. Furthermore, intelligent concurrency approaches with a low latency footprint can be used, so that scheduling impacts are minimised. From a protocol perspective, these reductions provide more time for encoding and decoding of blocks, or could even allow additional retransmission rounds on links with low delay.

Furthermore, precise bounds on the processing delay of the involved system components can increase the reliability of PRRT and reduce margins allocated for this size that is hard to predict with normal operating systems. Having these bounds, lost information can be detected easier, as overly delayed feedback due to a busy peer can be avoided. Furthermore, the retransmit timer now only needs to take channel variations into account, as processing delays are constant. Clock synchronisation and channel estimation require time-stamping of packets and communicating these values. This process can yield better results, as more precise time-stamps can be stored in the packets, again reducing the impact of system latency. As PRRT provides channel measurements to  applications, better estimates can improve application performance, i.e. leading to increased Quality-of-Control or Quality-of-Experience.

PRRT enables advanced applications in the area of CPS, but while it does not rely on many functions of the underlying layer, its operation is still limited by the transmission characteristics these layers can provide. In particular, this also includes the predictability of the operating system it runs on. While we assume that system level reliability is given by modern software development processes, the latency is an important area, where PRRT's performance can improve with optimisations on the system level. Consequently, latencies are reduced, bounded or even both.

\subsection{Timing Measurement Infrastructure}

When designing and optimising a protocol such as PRRT, it is imperative to precisely and thoroughly profile its performance, in particular regarding timing behaviour. This profiling has to be executed in a cross-layer fashion, taking communication and system aspects into account.

\subsubsection{Fine-grained Timing Measurements}

In general, profiling of system services faces two challenges. First, time-measurements must be accurate enough to allow profiling of relatively short code paths. Second, the run-time overhead of time measurement procedures should be minimal. Otherwise, the measurement itself could distort the run-time behaviour, and thus cause incorrect results. This evaluation therefore uses two interfaces to measure time, \texttt{clock\_gettime} and \texttt{rdtsc}.

The Linux system call \texttt{clock\_gettime}, given \texttt{CLOCK\_MONOTONIC} as clock identifier, returns the elapsed wall-clock time with up to nanosecond precision. However, this system call has a considerable overhead. On our evaluation platforms, two consecutive calls differ by circa $70ns$, which indicates relatively high run-time costs considering that \xlap also analysis the execution time of small protocol fragments.

In order to measure the latency of short code paths, the evaluation uses the  x86 instruction \texttt{rdtsc}. This instruction reads a hardware counter, which the CPU increments every processor cycle. This interface therefore enables measurements with approximately clock-cycle granularity, and with minimal overhead.

In order to actually measure execution times within the analyzed protocol, the evaluation combines the \texttt{clock\_gettime} and \texttt{rdtsc} interfaces. First, the coarse-grained \texttt{clock\_gettime} system call evaluates the execution time of relatively large code fragments, such as an end-user functions \texttt{send} and \texttt{receive}. Second, a \texttt{rdtsc} instruction is attached to every clock query, allowing to relate cycle counter values (``cycle-stamps'') to wall-clock values (``time-stamps''). Third, fine-grained timing measurements are performed using the \texttt{rdtsc} instruction, and by linear interpolation of cycle-stamps and time-stamps.

\subsubsection{Latency and Jitter Analysis}

An important goal of \xlap is the identification of the root causes of latency and jitter. To this end, precise information for each network packet is required. Therefore, \xlap provides a table data structure to store all time-stamps and cycle-stamps gathered during evaluation. In this table, the sequence number of the packet is used as row number, and the code location associated with the particular time-stamp or cycle-stamp serves as the column number. To avoid interference (\emph{false sharing}) between multiple worker threads, each time-stamp and cycle-stamp value is aligned to a cache-line.

Besides fast information disposal, the time-stamp table allows for precise profiling of individual packets. Since the association between packets and individual time-stamps is implicitly stored, each packet can be individually analysed for latency.

\subsubsection{Inter-Host Timing Measurements}

After the evaluation round, including all measurement packets, \xlap dumps the entire time-stamp table, which includes every single time-stamp and cycle-stamp, into a \texttt{csv} file. Thereby, sender and receiver each produce a data file, which \xlap aggregates \emph{post-experiment} to avoid interference. Since sequence numbers are equal on both endpoints, the aggregation reveals both end-to-end and cross-layer latency information.

Combining the timing datasets from multiple hosts demands for clock synchronisation, but this is only possible up to a certain extent~\cite{Mills:2012:PTP}. In consequence, the communication endpoints can possibly have slightly desynchronised clocks, which skew the link latency in our measurements. The evaluation in the following sections, however, focusses on processing delays of a reliable transport protocol rather than channel properties.

\section{Evaluation and Analysis}
\label{sec:evaluation}

Knowing the implementation details of a protocol, in our case PRRT, its timing behaviour can be empirically evaluated and analysed, using the lightweight time-stamping facilities of \xlap. The goal is to track root causes of latency and jitter.

\subsection{Methodology}

The \texttt{csv} files generated by \xlap include all captured packets, identified by their sequence number and packet type, as well as all related time- and cycle-stamps. Sender and receiver \texttt{csv} contain all columns, but have zeros for those stamps that are only taken on the other side. Furthermore, many time-stamps are 0, because only cycle-stamps are taken at these specific processing step.

The analysis begins with combining and completing the data set captured by \xlap. Firstly, data-frames are generated and filtered by the type of packet, so that source and redundancy packets are analysed independently. Second, the data-frames of both sending and receiving sides are joined, providing end-to-end traces indexed by sequence number, for any transmitted packet. Third, the processing durations on both sides are determined, using the time-stamps that are gathered upon entering and leaving the PRRT protocol layer, as well as the channel time:
\begin{align}
    \Delta T_{SenderTotal} &= T_{LinkTransmitEnd} - T_{PrrtSendStart}\\
    \Delta T_{ReceiverTotal} &= T_{PrrtDeliver} - T_{LinkReceive}\\
    \Delta T_{Channel} &= T_{LinkReceive} - T_{LinkTransmitEnd}
\end{align}

Channel characteristics are currently out of the scope of the analysis, since this paper focusses on processing delays in the endpoints. Therefore, receiver time-stamps are adjusted, by subtracting the channel time. Consequently, we consider delivery to be instant. Besides, we thus avoid problems with clock synchronisation.
\begin{align}
    \Delta T_{E2E} &= \Delta T_{SenderTotal} + \Delta T_{ReceiverTotal}
\end{align}

\begin{figure}
\begin{tikzpicture}

\tikzset{axis/.style={-latex,thick}}
\tikzset{both/.style={thick,draw=black,fill=blue!35,circle}}
\tikzset{line/.style={thick,draw=blue!45!black}}
\tikzset{tick/.style={thick}}
\tikzset{map/.style={dotted,thick}}

\def\xmax{6.5}
\def\ymax{2.2}
\def\tckl{0.2}

\def\min{0.25}
\def\max{0.8}
\def\mid{0.6}

\draw[axis] (0,0) -- ++(\xmax,0) node[below] {Cycles};
\draw[axis] (0,0) -- ++(0,\ymax) node[above] {Time};

\node[both] (N1) at ($(\min*\xmax,\min*\ymax)$) {};
\node[both] (N2) at ($(\max*\xmax,\max*\ymax)$) {};

%\draw[line] (0,0) -- ($(0.9*\xmax,0.9*\ymax)$);
%\draw[line] ($(\min*\xmax,\min*\ymax)$) -- ($(\max*\xmax,\max*\ymax)$);
\draw[line] (N1) -- (N2);

\draw[tick] ($(\min*\xmax,\tckl)$) -- ++($(0,-2*\tckl)$) node[below] {$C_{Start}$};
\draw[tick] ($(\max*\xmax,\tckl)$) -- ++($(0,-2*\tckl)$) node[below] {$C_{End}$};
\draw[tick] ($(\mid*\xmax,\tckl)$) -- ++($(0,-2*\tckl)$) node[below] {$C_{i}$};

\draw[tick] ($(\tckl,\min*\ymax)$) -- ++($(-2*\tckl,0)$) node[left] {$T_{Start}$};
\draw[tick] ($(\tckl,\max*\ymax)$) -- ++($(-2*\tckl,0)$) node[left] {$T_{End}$};

\draw ($(-1*\tckl,\mid*\ymax)$) node[left] {$T_{i}$};

\draw[map,-latex] ($(\mid*\xmax,0)$) -- ++($(0,\mid*\ymax)$);
\draw[map,-latex] ($(\mid*\xmax,\mid*\ymax)$) -- ($(0,\mid*\ymax)$);

\draw[map] ($(\min*\xmax,\tckl)$) -- (N1);
\draw[map] ($(\max*\xmax,\tckl)$) -- (N2);
\draw[map] ($(\tckl,\min*\ymax)$) -- (N1);
\draw[map] ($(\tckl,\max*\ymax)$) -- (N2);

\end{tikzpicture}
\caption{Sketch of the Time-stamp Reconstruction}
\label{fig:csts_conversion}
\end{figure}
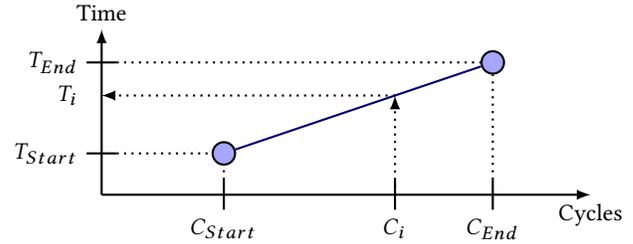

Missing time-stamps for processing steps where only cycle-stamps are taken are now reconstructed using a conversion sketched in Fig.~\ref{fig:csts_conversion}. On both hosts, \xlap measures the start and end time, and an accompanied cycle-stamp. Intermediate time-stamps, such as $T_{i}$, are reconstructed from the corresponding cycle-stamp $C_{i}$ using linear interpolation. Thereby, the slope of the graph corresponds to the processor frequency.

To foster jitter analysis, durations that relate to specific code blocks are determined using the previously recovered time-stamps. \xlap further provides utility functions that execute linear regressions on the data sets and generate histograms and scatter plots. This functionality proved useful for optimising the \xlap infrastructure itself, but can also be leveraged for detecting variations of delays over time, in particular to detect trends. Finally, functions for generating packet traces and jitter analysis are included, which are extensively discussed in Sec.~\ref{sec:results}.

Executing the analysis is done using the Python library \emph{Pandas}\footnote{\url{http://pandas.pydata.org/}} and plots are generated using \emph{matplotlib}\footnote{\url{http://matplotlib.org/}} inside a Python3 \emph{jupyter}\footnote{\url{http://jupyter.org/}} notebook. The previously mentioned raw data in form of \texttt{csv} files and the notebook are freely available and can be found online\footnote{\url{https://git.nt.uni-saarland.de/LARN/X-Lap}}.

\subsection{Experimental Setup}

The experiments were executed in different scenarios, on a single PC, a node pair, and a networking testbed. The environment using loop-back interfaces is useful for debugging \xlap, but the lack of network-related jitter is noticeable in the results. The following evaluation uses a PC pair and the testbed, which produce similar results. The testbed hosts have 8GB memory and 8 cores, ensuring that no resource limitation distorts the evaluation. It should nevertheless be noted that the test systems use Linux without any adjustments for real-time, hence the scheduler impacts performance.

\subsection{Results and Analysis}
\label{sec:results}

Following the goal to find root causes of latency and jitter, different evaluation procedures are included in \xlap. This allows to inspect individual packet traces but also detect correlations between series of packets.

\subsubsection{Packet Traces}
\begin{figure}
    \includegraphics[width=\columnwidth]{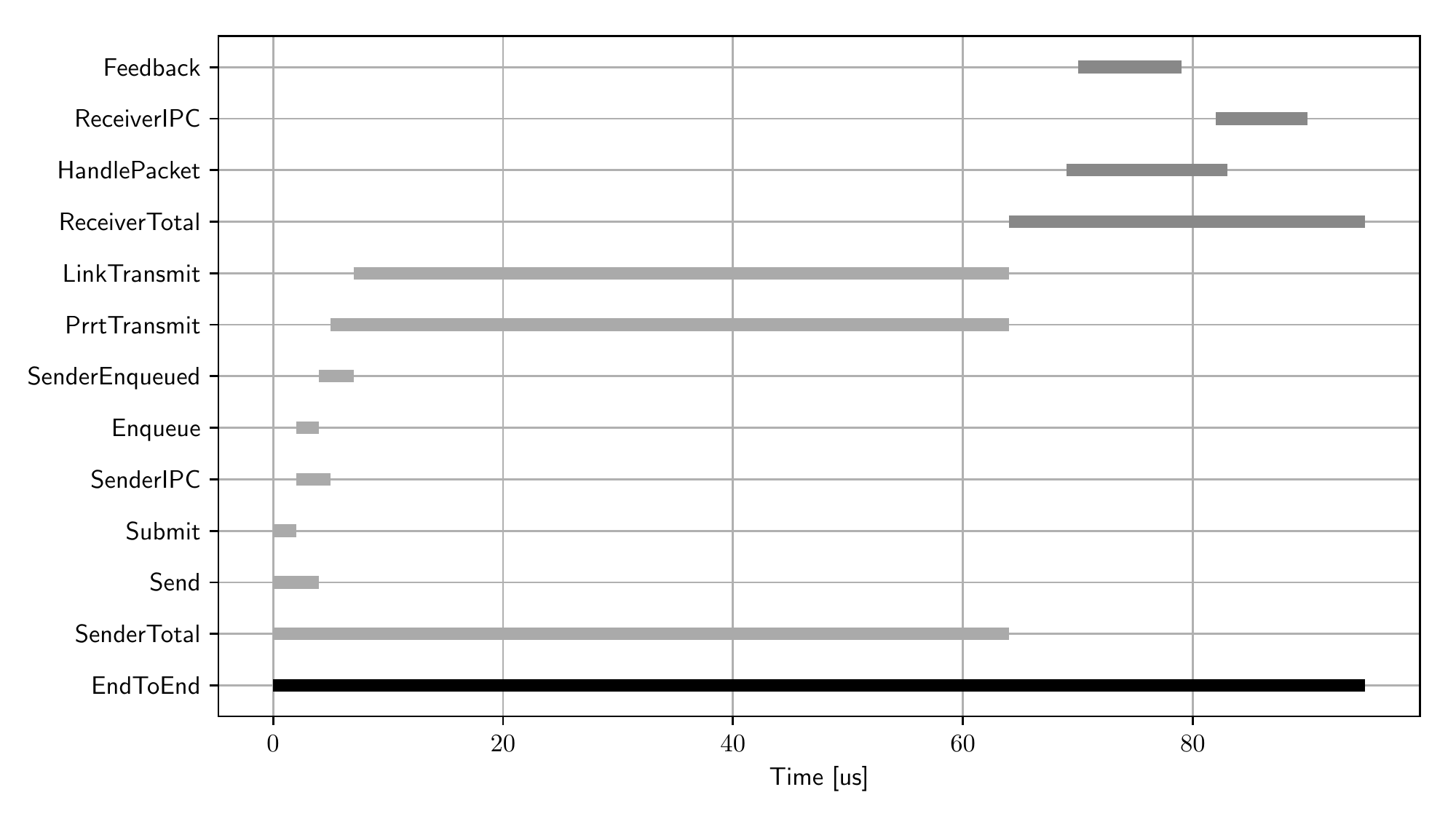}
    \caption{Detailed Trace of a Representative Packet}
    \label{fig:trace}
\end{figure}

By measuring and reconstructing time-stamps of the processing of a single packet, it is possible to provide a trace across sender and receiver, as is depicted in Fig.~\ref{fig:trace}. These traces reveal which processing phases overlap, indicating existing latency hiding, and which ones happen in succession. For a given packet, it is possible to compare the latency induced by individual steps. For instance, the depicted trace shows that sender latency dominates compared to receiver and that this is mainly due to UDP socket transmission costs.

\subsubsection{Trace Jitter}
\begin{figure}
    \includegraphics[width=\columnwidth]{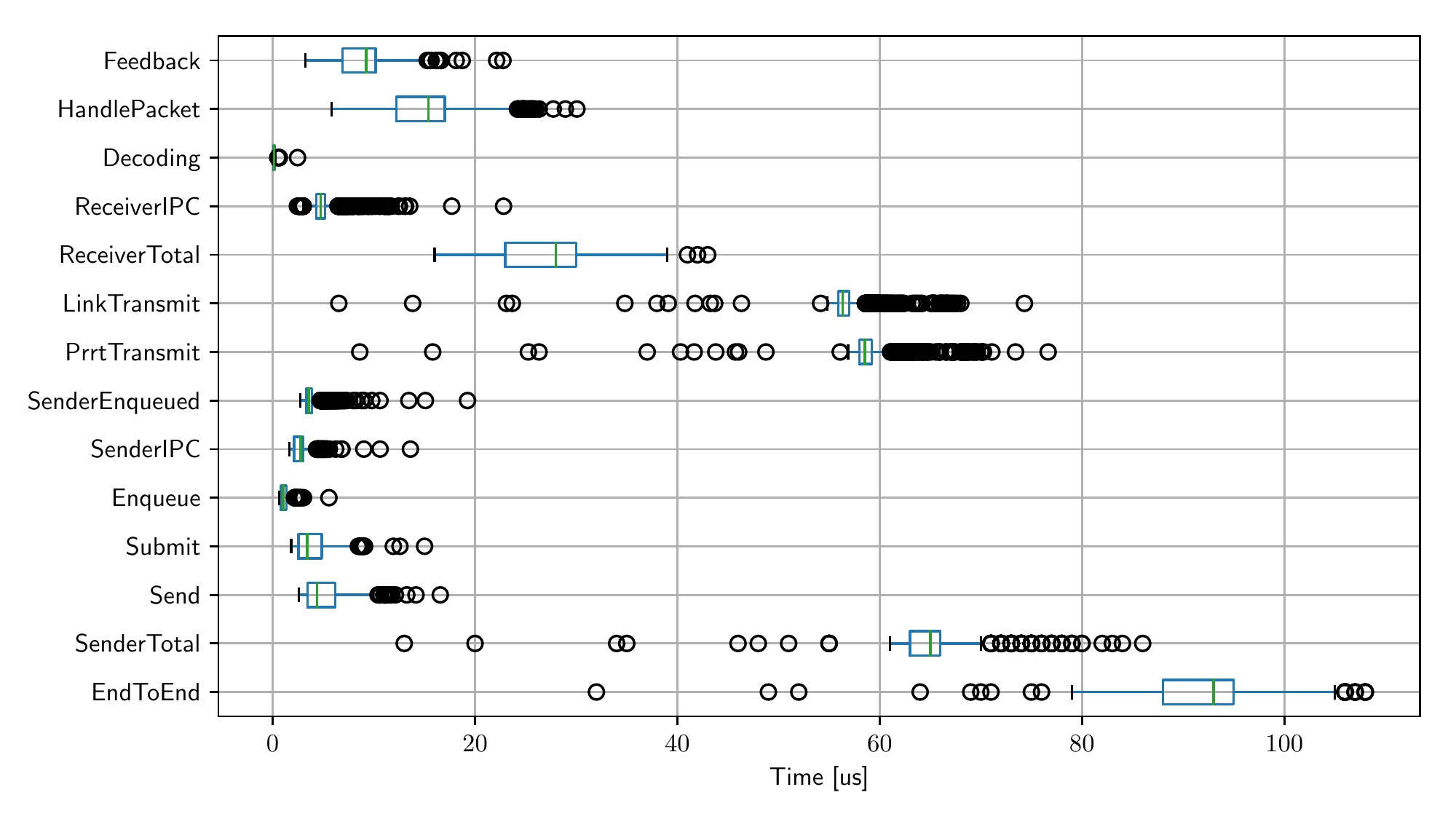}
    \caption{Overview of Packet Transmission Jitter}
    \label{fig:jitter}
\end{figure}

While the former approach allows to quantify latency, comparing the variations within individual processing steps reveal the sources of jitter. To this end, we extract outliers from the data set. We consider traces to be outliers regarding one parameter if the parameter is above the 75\,\% quantile plus 1.5 \emph{Inter-Quantile-Range} (IQR). The traces where the end-to-end time is considered as an outlier are further analysed in Sec.~\ref{subsec:outliers}, while this section focusses on values below this threshold.

The visualisation in Fig.~\ref{fig:jitter} uses box plots, where the median is marked as a green line and the 25\,\% and 75\,\% quantiles form the outer borders of the box. The whiskers indicate the most extreme value that is within 1.5 of the IQR and outliers are marked with circles. We can see that sender-sided times, in particular for packet transmission, are facing high jitter.

\subsubsection{Outliers}
\label{subsec:outliers}

\begin{figure}
    \includegraphics[width=\columnwidth]{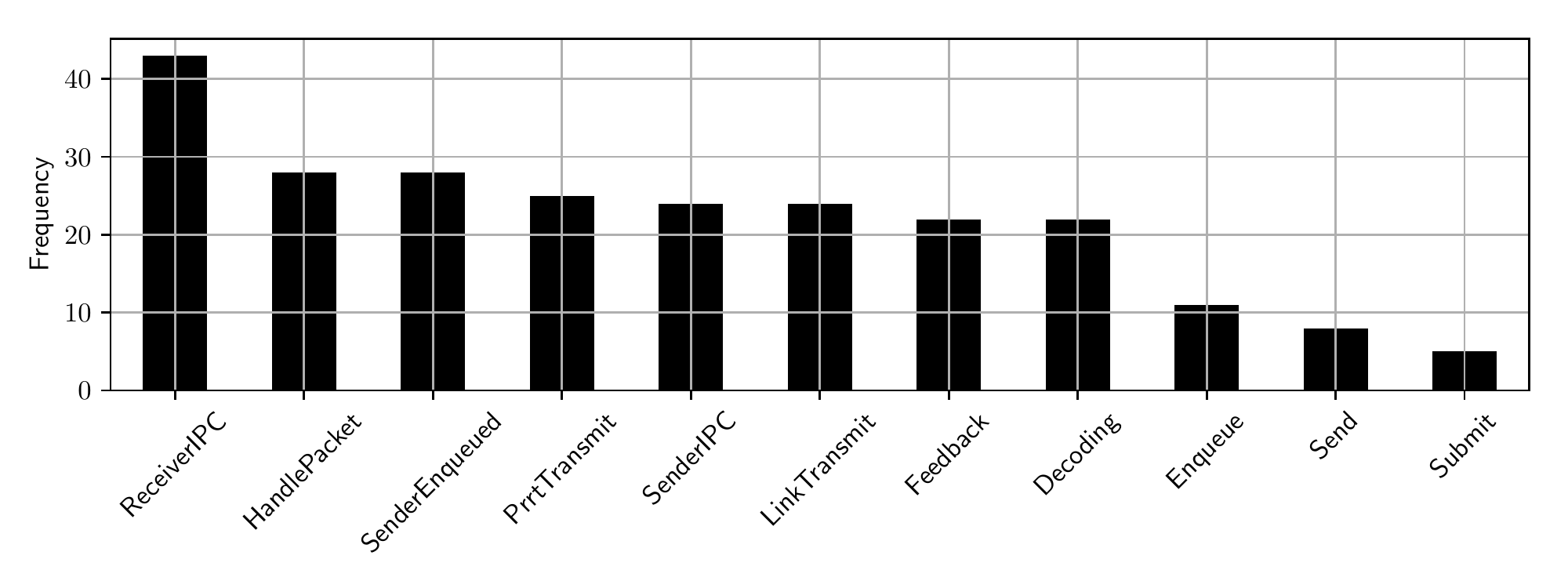}
    \caption{Jitter Causes (Threshold: 108.0us, 75 samples)}
    \label{fig:causes}
\end{figure}

The previously separated outliers regarding end-to-end time are now considered further, regarding the root cause of the increased delay. By concept, the end-to-end delay is the sum of multiple partial latencies. Therefore, we consider a protocol component a \emph{cause} of end-to-end jitter if the corresponding latency is also an outlier. Fig.~\ref{fig:causes} depicts the distribution of how often specific protocol parts cause end-to-end jitter. The result indicate that \emph{Inter-Process Communication} (IPC) on the receiver is often responsible for high end-to-end delay, and thus give an indication where outliers can be eliminated.

\subsubsection{Correlation}
\begin{figure*}
    \includegraphics[width=\textwidth]{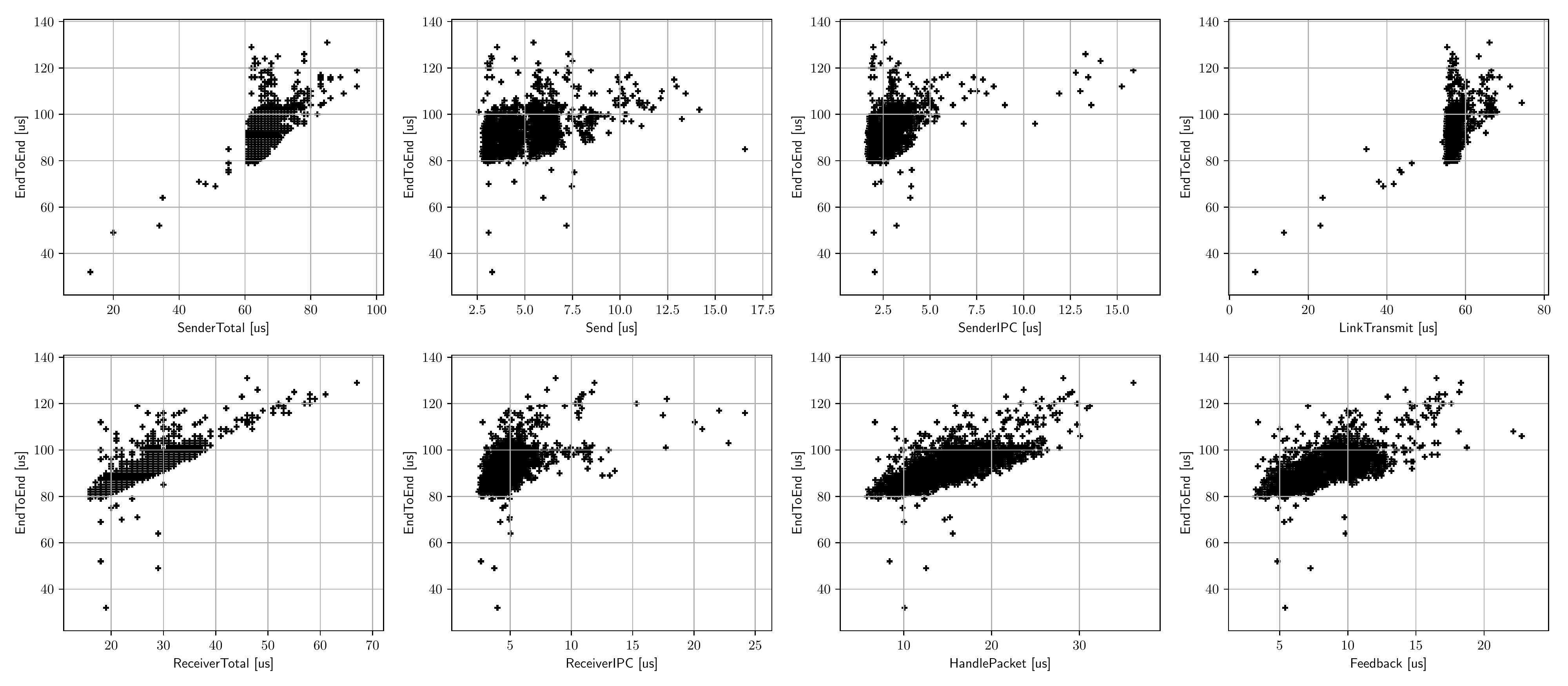}
    \caption{Correlation Between Individual Latencies and End-To-End Time}
    \label{fig:correlation}
\end{figure*}

Finally, in order to further identify potential causes for increased latencies, correlations between the end-to-end time and partial latencies are given in Fig.~\ref{fig:correlation}. The graphs show that the sender has a high base latency of $60us$, but its execution time correlates weakly to the end-to-end duration.
Instead, the graphs proves the impact of unusual receiver processing time on the overall performance. In particular, the latency caused by packet handling and feedback sending has a direct effect on the end-to-end time. Furthermore, the graphs show that high IPC latency co-occurs with high end-to-end latency.
Even though this analysis only reveals correlations, and no causal relations, it gives valuable insight how unpredictability in protocol parts corresponds to end-to-end jitter.

\section{Related Work}
\label{sec:relatedwork}

To the best of our knowledge, this paper is the first approach to provide cross-layer, inter-host timing analysis for real-time networking stacks. Previous work often has a focus on timing models, or it evaluates individual host systems without considering any networking components.

Schimmel et al.~\cite{Schimmel:IEC} compute the \emph{Worst-Case Execution Time} (WCET) to provide upper bounds on communication delay, and prove that application requirements (\emph{deadlines}) are met. The authors assume closed systems, which often holds for industrial control systems, so that there is no competing traffic and the channel remains static. These assumptions were not made when designing PRRT, because it significantly limits the areas in which the protocol can be implemented. Furthermore, \xlap can validate such timing models empirically for any communication protocol.

It has been noted by Liu et al.~\cite{Liu:2016} that in the area of \emph{Network-on-Chip} (NoC), which can be considered as a building block for CPS and CPN, there is a lack of investigations and design methodologies on schedulability. With real-time applications, this is a crucial trait that is by far more important than having a maximum throughput. This paper specifically deals with NoCs and their abstract representations, with the goal of optimising its performance with a specific approach. We follow a general approach that is applicable on any network, and is primarily used to identify potential spots for improvement. Eventually, these tools should be used to provide a \emph{Worst-Case Traversal Time (WCTT)}~\cite{Ferrandiz:2009}, to enable networks of real-time applications, when the hardware is known.

The influence of seemingly minor operating system functions towards network~\cite{mraz:1994:sc} and application~\cite{tsafrir:2005:ics,beckman:2006:iccc} performance has been summarised under the term \emph{OS noise}. A well-known source of operating system noise are hardware interrupts. If such an interrupt, or other system activity, slows down a single process, all depending processes have to wait. In consequence, minor delays can accumulate to a significant performance and predictability problem. This issue occurs at network protocols where protocol components have inherent data dependencies. For \emph{High Performance Computing} (HPC) systems, a typical solution to OS noise is the use of \emph{lightweight kernels} that improve timing predictability by omitting unnecessary functionality.

Barroso et al.~\cite{barroso:2017:cacm} argue that various delays in the scale of microseconds accumulate and harm network performance significantly. While their work focusses on throughput-oriented data-center networking, their key observations are aligned with the results of this paper. The authors propose hardware support for latency hiding.

\section{Conclusion}
\label{sec:conclusion}

Real-time networks and CPNs need a reliable distributed tool-chain for latency and jitter analysis. In this paper, we have proposed \xlap, a timing analysis tool particularly tailored to the needs of real-time communication. Furthermore, this paper analyses the reliable real-time communication protocol PRRT. Our results show that operating system primitives, especially IPC, can have a significant impact on latency and jitter. These insights are going to be used for improving PRRT, while the development of \xlap allows to analyse other protocols such as TCP on Linux. We therefore propose a co-design approach that treats the application, operating systems and network protocols as a unit.

\begin{acks}
    The work is supported by the \grantsponsor{DFG}{German Research Foundation~(DFG)}{TODO} as part of SPP 1914 ``Cyber-Physical Networking'' under grants \grantnum{DFG}{HE~2584/4-1} and \grantnum{DFG}{SCHR~603/15-1}.
\end{acks}

\bibliographystyle{ACM-Reference-Format}
\bibliography{ms}

\end{document}